\documentclass[conference]{IEEEtran}
\IEEEoverridecommandlockouts
\usepackage{cite}
\usepackage{amsmath,amssymb,amsfonts,amsthm}
\usepackage{algorithmic}
\usepackage{graphicx}
\graphicspath{{figs/}}
\usepackage{textcomp}
\usepackage[]{authblk}

\usepackage{xcolor}
\definecolor{LinkColor}{cmyk}{1,0.50,0,0.1}
\usepackage[unicode,colorlinks=true,allcolors=LinkColor]{hyperref}

\usepackage{comment}
\def\BibTeX{{\rm B\kern-.05em{\sc i\kern-.025em b}\kern-.08em
    T\kern-.1667em\lower.7ex\hbox{E}\kern-.125emX}}

\usepackage{graphicx}	
\usepackage{amsmath}	
\usepackage{comment}
\usepackage{xspace} 
\usepackage{mathtools}
\usepackage{stmaryrd}
\usepackage{float}
\usepackage{bm}
\usepackage{pdflscape}
\usepackage{pdfpages}
\usepackage{empheq}
\usepackage{mathrsfs}
\usepackage[font=small,skip=4pt]{caption}
\DeclareFontFamily{U}{mathx}{}
\DeclareFontShape{U}{mathx}{m}{n}{<-> mathx10}{}
\DeclareSymbolFont{mathx}{U}{mathx}{m}{n}
\DeclareMathAccent{\widecheck}{0}{mathx}{"71}

\usepackage{multirow} \usepackage{threeparttable}
\usepackage{booktabs} \usepackage{stfloats}

\newcommand*{\V}[1]{\boldsymbol{#1}}   
\newcommand*{\M}[1]{\mathbf{#1}}       
\newcommand*{\TransposeLetter}{\hspace*{-.25ex}\top\hspace*{-.25ex}}
\newcommand*{\T}{^{\TransposeLetter}} 
\DeclareFontFamily{U}{mathx}{\hyphenchar\font45}
\DeclareFontShape{U}{mathx}{m}{n}{<-> mathx10}{}
\DeclareSymbolFont{mathx}{U}{mathx}{m}{n}

\DeclarePairedDelimiterX{\Paren}[1]{(}{)}{#1}
\DeclarePairedDelimiterX{\Brace}[1]{\{}{\}}{#1}
\DeclarePairedDelimiterX{\Brack}[1]{[}{]}{#1}
\DeclarePairedDelimiterX{\Abs}[1]{\rvert}{\lvert}{#1}
\DeclarePairedDelimiterX{\Norm}[1]{\lVert}{\rVert}{#1}
\DeclarePairedDelimiterX{\Avg}[1]{\langle}{\rangle}{#1}
\DeclarePairedDelimiterX{\Round}[1]{\lfloor}{\rceil}{#1}
\DeclarePairedDelimiterX{\Floor}[1]{\lfloor}{\rfloor}{#1}
\DeclarePairedDelimiterX{\Ceil}[1]{\lceil}{\rceil}{#1}
\DeclarePairedDelimiterX{\Inner}[2]{\langle}{\rangle}{#1,#2}
\newcommand*{\delimsize}{}  
\newcommand*{\Given}{\:\delimsize\vert\:} 

\DeclareMathOperator*{\argmin}{argmin}

\DeclareMathOperator{\Trace}{tr}

\DeclarePairedDelimiterXPP{\Expect}[1]{\mathbb{E}}(){}{#1}

\newcommand*{\estim}[1]{\widehat{#1}}

\newcommand*{\Set}[1]{\mathbb{#1}}

\usepackage[squaren,Gray]{SIunits}

\newcommand*{\CoLogLikelihood}{\mathcal{L}}

\newcommand*{\Hyp}[1]{\ensuremath{\mathscr{H}_{#1}}}
\newcommand*{\TestSup}{\overset{\Hyp1}{\underset{\Hyp0}{\gtrless}}}

\newcommand*{\Iter}[1]{^{[#1]}}

\newcommand*{\Range}[2]{#1{:}#2}

\newcommand*{\bydef}{\stackrel{{\scriptscriptstyle \text{def}}}{=}}

\usepackage{numprint}

\makeatletter
\def\widebreve{\mathpalette\wide@breve}
\def\wide@breve#1#2{\sbox\z@{$#1#2$}%
     \mathop{\vbox{\m@th\ialign{##\crcr
\kern0.08em\brevefill#1{0.8\wd\z@}\crcr\noalign{\nointerlineskip}%
                    $\hss#1#2\hss$\crcr}}}\limits}
\def\brevefill#1#2{$\m@th\sbox\tw@{$#1($}%
  \hss\resizebox{#2}{\wd\tw@}{\rotatebox[origin=c]{90}{\upshape(}}\hss$}
\makeatletter

\begin{document}

\title{Learning a model of non-stationary nuisances with non-local correlations for object detection
{}
\thanks{This work was supported by the ANR under the
France 2030 program (PEPR Origins, ANR-22-EXOR0016) and
by the Action Spécifique Haute Resolution Angulaire (ASHRA) of
CNRS/INSU co-funded by CNES.}
}

\author{Olivier Flasseur$^1$, Éric Thiébaut$^1$, Loïc Denis$^2$ and, Samuel Thé$^3$.}
\affil{\small
  $^1$Universite Claude Bernard Lyon 1, CRAL UMR 5574, ENS de Lyon, CNRS, Villeurbanne, F-69622, France \protect\\
  $^2$Université Jean Monnet Saint-Etienne, CNRS, IOGS, Laboratoire Hubert Curien UMR 5516, Saint-Etienne, F-42023, France \protect\\
  $^3$Haystack Observatory, Massachusetts Institute of Technology, Westford, MA}

\maketitle

\begin{abstract}
  We tackle the problem of detecting faint objects in multivariate data corrupted by a
  strong and spatially correlated nuisance component in the low-sample regime. Standard
  covariance-based models often yield unstable or ill-conditioned estimators of the nuisance
  statistics, severely degrading detection performance. We propose a data-driven approach
  based on sparse precision matrix modeling, where the precision matrix is approximated
  through a pivoted sparse Cholesky factorization whose non-zeros coefficients are learned
  by maximum likelihood. This yields a low-complexity yet expressive statistical model of
  the nuisance. Building on this model, we derive a generalized likelihood ratio test for
  faint source detection enabling direct control of the false alarm rate. The approach is
  evaluated on an astrophysical application: the detection of exoplanets by high-contrast
  imaging, where it achieves state-of-the-art performance.
\end{abstract}


\vspace*{-1mm}
\section{Introduction}
\label{sec:introduction}

\noindent Detecting faint objects in multivariate data corrupted by a structured
nuisance component is a central problem in various
applications, including biomedical imaging, remote sensing, and astronomy.
Many detection methods rely on second-order statistics to model
the nuisance, but in practice the number of available samples is often limited.
This low-sample, high-dimensional regime leads to unstable or singular covariance
estimators, severely degrading maximum likelihood-based inference and detection
performance. Achieving a bias–variance trade-off is thus a key
challenge in such inverse problems exploiting measurement statistics.

A possible solution is to calibrate the nuisance statistics using
additional data independent of the observations of interest. However, this is
often impractical, as the statistics can vary significantly across datasets
(e.g., due to changing conditions of observation over time in radar imaging or
astronomy).
Self-calibrating approaches address this challenge through covariance regularization.
These include diagonal loading and shrinkage methods \cite{chen2010shrinkage,ollila2019optimal}, which stabilize the
estimator by combining the sample covariance with another estimator of lower variance, as well as
approaches that impose predefined covariance structures, such as block-diagonal
or banded models, when long-range correlations are assumed negligible \cite{bickel2008regularized}. Hybrid
strategies combining structural assumptions and nonlinear shrinkage have also been
proposed \cite{ledoit2020analytical}. While these methods improve numerical stability, they rely by
construction on strong prior assumptions about the covariance structure.
A more flexible alternative consists in directly modeling the inverse covariance
matrix, or precision matrix, which often admits sparse representations \cite{higham2009cholesky}. In
particular, sparse Cholesky factorizations provide a structured and
computationally efficient way to approximate precision matrices while
guaranteeing positive semidefiniteness \cite{schafer2021sparse}. However, defining the sparse structure,
learning the corresponding non-zero coefficients in a fully data-driven manner,
and integrating such models into an optimal detection framework remain challenging.

\textit{Our contributions.}
We propose a low-complexity, data-driven detection framework based on sparse precision matrix modeling
of the nuisance component. The precision matrix is approximated by a pivoted
sparse Cholesky factorization learned from the data by maximum likelihood,
enabling stable estimation in the low-sample regime while capturing long-range
(spatial) correlations (Sect.~\ref{sec:learning_nuisance}). Building on this model,
we derive a detection test with a constant false alarm
rate (Sect.~\ref{sec:detection}). The model complexity is selected
automatically via cross-validation (Sect.~\ref{sec:loocv}). We demonstrate the effectiveness of the proposed
approach for exoplanet detection
(Sect.~\ref{sec:results}).

\section{Low-Complexity Precision Matrix Model}
\label{sec:learning_nuisance}

\subsection{Problem statement}
\label{subsec:problem_statement}

\noindent We consider the general case of $N$ observations, where the $n$-th data sample
$\V y_n \in \mathbb{R}^P$ follows the forward model:
\begin{equation}
	\V y_n = \V f_n(\V x) + \V s_n\,, \qquad (\forall n \in \Range{1}{N})\,,
	\label{eq:direct_model}
\end{equation}
where $\V f_n{:}\ \mathbb{R}^L \rightarrow \mathbb{R}^P$ denotes a known (possibly
nonlinear) mapping of the unknown parameters $\V x \in \mathbb{R}^L$, and ${\V s}_n$ is a
nuisance component with ${\V s}_n \gg \V f_n(\V x)$. The mapping $\V f_n$ depends on the
considered task (object detection, image reconstruction, etc.). We also assume $P \gg N$,
thus defining a low-sample regime.

\subsection{Model of the nuisance component}
\label{subsec:mle_problem}

\noindent We assume that the nuisance terms $\lbrace \V s_n \rbrace_{n\in\Range{1}{N}}$ are
random variables that are mutually independent (for $n' \neq n$) but follow identical
Gaussian distribution:
\begin{equation}
  \V s_n \sim \mathcal{N}(\V \mu,\, \M \Lambda^{-1})\,, \qquad (\forall n \in \Range{1}{N})\,,
  \label{eq:nuisance_distribution}
\end{equation}
with $\V \mu \bydef \Expect{\V s_n}$ the expectation of the nuisance terms and $\M \Lambda$
their precision matrix. We denote by $\M \Omega = \lbrace \V \mu,\, \M \Lambda \rbrace$ the
nuisance statistics. Under forward model~\eqref{eq:direct_model}, the probability density
function of the $n$-th nuisance term is given by:
\begin{equation}
  \text{pdf}\Paren*{\V s_n \Given \M \Omega} =
  \Abs*{\frac{\M \Lambda}{2\,\pi}}^{1/2}\,\exp\Paren*{
    - \frac{1}{2}\,\Norm[\big]{
      \V y_n - \V f_n(\V x) - \V \mu
    }^2_{\M \Lambda}
  }\,,
  \nonumber
  \label{eq:pdf-nuisance}
\end{equation}
\noindent where $\Abs{\M \Lambda}$ is the determinant of $\M \Lambda$, and
$\Norm{\V v}^2_{\M \Lambda} \bydef \V v\T\cdot\M \Lambda \cdot\V v$ is the squared
Mahalanobis norm.
Under these assumptions, the co-log-likelihood of the data sequence
$\lbrace \V y_n \rbrace_{n\in\Range{1}{N}}$ given $\V x$ and $\M \Omega$ writes:
\begin{equation}
  \CoLogLikelihood(\V x, \M \Omega) =
  \frac{1}{2}\,\sum_{n=1}^{N}
  \Norm[\big]{\V y_n - \V f_n(\V x) - \V \mu}^2_{\M \Lambda}
  - \frac{N}{2}\,\log\Paren{\Abs{\M \Lambda}}
  \label{eq:co-log-likelihood}
\end{equation}
up to an irrelevant additive constant. We address in
Sects.~\ref{subsec:mle_nuisance_statistics}-\ref{sec:detection} how to accurately estimate
all the unknown parameters, $\V x$ and $\M \Omega$, from the finite set of $N$ data samples.

\subsection{Maximum likelihood nuisance statistics}
\label{subsec:mle_nuisance_statistics}

\noindent Under the models of
Sects.~\ref{subsec:problem_statement}-\ref{subsec:mle_problem}, the maximum likelihood
estimators of the statistics $\M \Omega$ knowing $\V x$ reduce to the sample estimators:
\begin{equation}
  \estim{\V \mu}(\V x) = \argmin_{\V \mu} \CoLogLikelihood(\V x, \M \Omega)
  = \frac{1}{N}\sum\nolimits_{n=1}^{N} \Paren{\V y_n - \V f_n(\V x)}\,,
  \label{eq:sample_mean}
\end{equation}
\begin{equation}
  \estim{\M \Lambda}(\V x) = \argmin_{\M \Lambda} \CoLogLikelihood(\V x, \M \Omega)
  = \Big(\underbrace{\frac{1}{N}\sum\limits_{n=1}^{N}\V r_n(\V x) \, \V r_n(\V x)\T}_{=\:\estim{\M C} (\V x)}\Big)^{-1}
  \label{eq:sample_precision}
\end{equation}
with the residuals $\lbrace \V r_n(\V x) = \V y_n - \V f_n (\V x) - \V \mu \rbrace_{n\in\Range{1}{N}}$.
Since the rank\footnote{When there is no ambiguity, we omit the dependence on $\V x$ for
  readability.} of $\estim{\M C}$ is at most $N$, a consistent covariance estimate requires
$N \gg P$. In the considered low-sample regime ($P \gg N$), $\estim{\M C}$ is therefore
singular, preventing direct estimation of the precision matrix $\estim{\M \Lambda}$ by
inversion of $\estim{\M C}$. To cope with this issue, we propose to use a matrix with a
lower complexity such as the matrices of \textit{A family of Sparse Approximations for
  whitening and Preconditioning} (ASAP, \cite{thiebaut2022beyond}) as an estimator of the
precision matrix. We thus approximate $\M \Lambda$ by:
\begin{equation}
  \M \Lambda \approx \M S\T \cdot \M S = \M P\T \cdot \M R\T \cdot \M R \cdot \M P  \,,
  \label{eq:precision-factorization}
\end{equation}
where $\M S$ is the composition of a sparse lower triangular matrix $\M R$ and of a
permutation matrix $\M P$, with $\M P\T = \M P^{-1}$. The symmetric and non-negative
semi-definite precision matrix $\M \Lambda$ defined in
Eq.~\eqref{eq:precision-factorization} can be seen as a pivoted sparse Cholesky
factorization \cite{higham2009cholesky}.

\subsection{Building the sparse structure and learning its coefficients}
\label{subsec:building_learning_sparse_model}

\noindent We decompose the construction of the precision matrix estimator $\M \Lambda$ in
\eqref{eq:precision-factorization} into two steps: (i) definition of the sparse structure,
and (ii) estimation of the associated coefficients.

\textit{Construction of the sparse structure}: The structure of $\M S$ is defined by (i) the
ordering of the \emph{nodes} (e.g., the pixels) and (ii), for each node, the set
of connected preceding nodes.
The ordering is specified by the permutation matrix $\M P$, while the connectivity is
defined by:
\begin{equation}
  \mathbb{S}_i = \lbrace j \in \Range{1}{i} \, | \, (i,\,j) \in \mathbb{S} \rbrace\,, \quad (\forall i \in \Range{1}{P})\,,
  \label{eq:set_Si}
\end{equation}
where $\mathbb{S}_i$ denotes the set of nodes connected to node $i$, and $\mathbb{S}$ is the
sparsity pattern of $\M R$, that is the set of index pairs $(i,j)$ of the structural
non-zeros of $\M R$. The model complexity $M$ is defined as the mean number of structural
non-zeros per row of $\M S$ and thus set the sparsity level of the model statistics.
There is a variety of choices to define an ordered sparse structure $\mathbb{S}$ that differ
by the sampling considered. A first solution is to assume $\M P = \M I_P$ (i.e., no
permutation) and to select, in \emph{lexicographic} order, for each node $i$, the previous
nodes that are closest to the node $i$:
\begin{equation}
  \Set S_i^{\text{lex}} = \lbrace j \in \Range{1}{i} \, | \, j \le i ,\, \mathcal{D}(i,j) \le \rho\rbrace\,,
  \label{eq:set_Si_lexicographic}
\end{equation}

where $\mathcal{D}(i,j)$ is a distance (e.g., Euclidean) between nodes $i$ and $j$, and
$\rho \ge 0$ controls the sparsity that is the number $M_i = |\mathbb{S}_i|$ of nodes
connected to the $i$-th one. A finer strategy uses an ordered \emph{random} sparse structure
based on a random permutation and in which a node $i$ is simply connected to the $M_i$
nearest preceding nodes. A related \emph{multi-scale} structure provides a more homogeneous
spatial coverage by constructing the permutation so that the density of the $n$ first nodes
is approximately uniform whatever $n \in \Range{1}{N}$.

\textit{Learning the sparse coefficients:} For a fixed sparsity pattern $\Set S$, the matrix
$\M S$ can be learned from the sample covariance \eqref{eq:sample_precision} in the maximum
likelihood sense as:
\begin{align}
  \estim{\M S}(\V x) &=
  \argmin_{\M S \in \mathcal{S}} \CoLogLikelihood(\V x, \M \Omega)
  \quad\text{s.t.}\quad \M \Lambda = \M S\T \cdot \M S,\\
  &= \argmin_{\M S \in \mathcal{S}} \Trace( \M S \, \estim{\M C} \, \M S\T) - \log (|\M S \T\, \M S |_+ )\,,
  \label{eq:best-ASAP-factor}
\end{align}
where $\mathcal{S}$ denotes the set of pivoted sparse Cholesky factors with given permutation $\M P$
and sparsity pattern $\Set S$.
Problem~\eqref{eq:best-ASAP-factor} can also be interpreted as the minimization of the
Kullback--Leibler divergence between two centered Gaussian distributions, one
with covariance $\estim{\M C}$ and the other with covariance $(\M S \T \, \M S)^{-1}$.
Given the sparse structure $\Set S$, Problem~\eqref{eq:best-ASAP-factor} has a closed-form solution for
the structural non-zero of $\M R$ \cite{thiebaut2022beyond}:
\begin{equation}
	\M R_{i,\mathbb{S}_i} = \frac{\pm \left( \big( \mathring{\M C}_{\Set S_i, \Set S_i} \big)^{-1} \right)_{M_i,1:M_i}}{\sqrt{\left( \big( \mathring{\M C}_{\Set S_i, \Set S_i} \big)^{-1} \right)_{M_i,M_i}}}\,, \quad (\forall i \in \Range{1}{P})\,,
	\label{eq:sparse_coeffs}
\end{equation}
where $\mathring{\M C}(\V x) = (1/N) \sum\nolimits_{n=1}^N (\M P \, \V r_n (\V x)) \cdot (\M P \, \V r_n (\V x))\T$ denotes the sample covariance of the permuted residuals, and $\mathring{\M C}_{\mathbb{S}_i,\mathbb{S}_i} \in \mathbb{R}^{M_i \times M_i}$ its restriction to the set $\mathbb{S}_i$ of column indices of the structural non-zeros of the $i$-th row of the sparse factor $\M R$ (also corresponding to the nodes directly connected to node $i$ in the sparse structure). Equation~\eqref{eq:sparse_coeffs} assumes that index $i$ is the last entry of $\mathbb{S}_i$, hence, at position $M_i$ in this set of node indices. The sign ambiguity
in~\eqref{eq:sparse_coeffs} does not affect the resulting approximation of
$\M \Lambda$ in~\eqref{eq:precision-factorization}.


\section{Detection of Faint Objects}
\label{sec:detection}


\noindent The statistical model of the nuisance introduced in
Sect.~\ref{subsec:mle_nuisance_statistics} is general and independent of the specific form
of the forward model $\V f_n(\V x)$ in Eq.~\eqref{eq:direct_model}. We now specialize this
model to the detection of objects parametrized by
$\V x = \lbrace \alpha,\, \V \theta \rbrace$, with $\alpha$ the object brightness and
$\V \theta$ its 2-dimensional position. In this setting, the forward model writes
$\V f_n(\V x) = \alpha\,\V p_n(\V \theta)$ with $\V p_n(\V \theta)$ the (known) footprint in
the image $\V y_n$ of an object at position $\V\theta$.

\subsection{Maximum likelihood brightness and uncertainty}
\label{subsec:mle_alpha}

\noindent The maximum likelihood estimator of the object brightness $\alpha$ knowing its
position $\V \theta$ and the statistics $\M \Omega$ is:
\begin{equation}
  \estim{\alpha}(\V \theta, \M \Omega)
  = \argmin_{\alpha \ge 0} \CoLogLikelihood(\alpha,\V \theta,\M \Omega)
  =\frac{\max\Paren{b(\V \theta, \M \Omega),0}}{a(\V \theta, \M \Omega)}
   \label{eq:alpha_hat}
\end{equation}
with:
\begin{align}
  \label{eq:a(θ)}
  a(\V \theta, \M \Omega) &= \sum\nolimits_{n=1}^{N}
  \Norm*{\V p_n(\V \theta)}_{\M \Lambda}^2 \, , \\
  \label{eq:b(θ)}
  b(\V \theta, \M \Omega) &= \sum\nolimits_{n=1}^{N}
  \V p_n(\V \theta)\T \cdot \M \Lambda \cdot \Paren{\V y_n - \V \mu} \, .
\end{align}
The statistic $b$ corresponds to a cross-correlation term, while $a$ is a normalization factor.
They provide sufficient statistics for the
estimation problem.
Indeed, under the Gaussian assumption with precision matrix $\M\Lambda$,
$\text{Var} [ b(\V \theta, \M \Omega) ] \approx \V p_n(\V \theta)\T \cdot \M \Lambda \cdot \V p_n(\V \theta) = a(\V \theta, \M \Omega)$.
It follows that the estimator of the flux standard-deviation is:
\begin{equation}
	\estim{\sigma}_\alpha (\V \theta, \M \Omega) = 1 / \sqrt{a (\V \theta, \M \Omega)}\,.
	\label{eq:sigma_hat}
\end{equation}

\subsection{Detection criterion}
\label{subsec:detection_criterion}

\noindent Assuming that the nuisance statistics $\M \Omega$ are known, detection is
performed by the generalized likelihood ratio test (GLRT):
\begin{equation}
  \Gamma(\V \theta, \M \Omega) = \frac{b(\V \theta, \M \Omega)}{\sqrt{a(\V \theta, \M \Omega)}}
  = \frac{\estim{\alpha}(\V \theta, \M \Omega)}{\estim{\sigma}(\V \theta, \M \Omega)}
  \TestSup \tau\,.
  \label{eq:detection_criterion}
\end{equation}
Under the null hypothesis~$\Hyp0$ and the assumed multivariate Gaussian nuisance model,
$\Gamma(\V \theta, \M \Omega)$ follows a standard normal distribution $\mathcal{N}(0,1)$,
independently of $\V\theta$. A single threshold $\tau > 0$ thus ensures a constant false
alarm rate (CFAR) detector and yields a natural interpretation of the detection criterion as
a signal-to-noise ratio (SNR). Detection is performed by exhaustively evaluating
\eqref{eq:detection_criterion} over a discretized (possibly subpixel) grid of candidate
positions. Since the sought sources are usually faint, nuisance statistics $\M\Omega$ are
estimated under $\Hyp0$ by assuming $\alpha=0$. After detection, the source flux and
position are locally refined by jointly re-estimating the nuisance statistics and source
parameters through a fixed-point iterative scheme, alternating closed-form flux updates
\eqref{eq:alpha_hat} and continuous position optimization maximizing
\eqref{eq:detection_criterion}.


\section{Unsupervised Model Complexity Selection}
\label{sec:loocv}

\noindent The performance of the proposed detector critically depends on the nuisance model
complexity $M$, which governs a bias–variance trade-off. If $M$ is too small, the model
fails to capture correlations in the data. If $M$ is too large, the model may overfit the
finite sample set and partially absorb the signal of interest into the estimated nuisance
statistics $\estim{\M\Omega}$. To set $M$ in an unsupervised manner, we assess how well the
nuisance model predicts unseen data rather than how well it fits the training samples. We
thus select $M$ by maximizing a leave-one-out cross-validated predictive log-likelihood
(LOOCV), defined as the sum of the co-log-likelihoods of each sample $\V y_n$ evaluated
under a model learned from all remaining samples.
\begin{equation}
  \mathcal{C}(M) = \frac{1}{2}\sum\limits_{n=1}^{N} \Norm[\big]{\V y_n - \estim{\V \mu}^{[n]} }^2_{\estim{\M \Lambda}^{[n]}} - \log\Paren{\Abs{\estim{\M \Lambda}^{[n]}}_+} \,,
  \label{eq:loocv}
\end{equation}
where $\estim{\M\Omega}^{[n]}= \lbrace \estim{\V\mu}^{[n]},\estim{\M\Lambda}^{[n]} \rbrace$ are the nuisance
statistics learned with model complexity $M$ from the training set
$\chi\Iter{n}= \lbrace\V y_{n'} \mid n'\in\Range{1}{N}, n' \not= n \rbrace$, i.e.,
excluding sample $\V y_n$. The LOOCV criterion $\mathcal{C}(M)$ therefore
quantifies how well a model of complexity $M$ generalizes to unseen data.
To avoid contamination by real (brighter) sources, $\mathcal{C}(M)$ must
be computed in an unbiased manner: pixels potentially affected by sources are
masked when learning and/or evaluating the nuisance statistics
$\estim{\M\Omega}\Iter{n}$. In practice, a preliminary detection map $\V\Gamma$ is
computed using a low-complexity model ($M=2$ or $3$) and thresholded at a high
level $\tau$ to conservatively identify and mask brighter sources, which are then
treated as invalid data in the LOOCV procedure.
The optimal model complexity $\estim{M}_{\text{LOOCV}}$ and the corresponding best sparse structure
(\textit{lexicographic}, \textit{random}, or \textit{multiscale}) are finally selected as the minimizers of
the LOOCV criterion.

\section{Results on Astronomical Data}
\label{sec:results}

\subsection{Considered modality: direct imaging at high-contrast}
\label{subsec:hci_principle}

\noindent We consider high-contrast imaging \cite{pueyo2018direct}, which aims at detecting
exoplanets appearing as unresolved point sources around an observed star. The data were
acquired with the SPHERE instrument at the Very Large Telescope, combining adaptive optics
and a coronagraph mask to suppress part of the starlight.
This yields spatio-temporal data cubes
$\lbrace \V y_n\rbrace_{n\in\Range{1}{N}} \in \mathbb{R}^{N\times P}$ with typically
$N \approx 50$ frames of $P \approx 10^6$ pixels.
Detection is challenging due to a strong, spatially correlated nuisance caused by residual
stellar leakage and optical aberrations, with exoplanet-to-star flux ratio (i.e.,
\emph{contrast}) of order $10^{-5}$ to $10^{-6}$ before optical correction and still below
$10^{-3}$ in the recorded images. This nuisance is both non-stationary and correlated at
multiple spatial scales (local speckles + extended structures). During the acquisition,
astrophysical sources follow a known rotation induced by Earth’s rotation, while the
nuisance remains quasi-static. In the forward model, $\V p_n(\V \theta)$ thus accounts for
this rotation and represents the spatial signature in the $n$-th frame of a source located
at the 2D sky position $\V \theta$.
\begin{figure}[t!]
	\centering
	\includegraphics[width=0.5\textwidth]{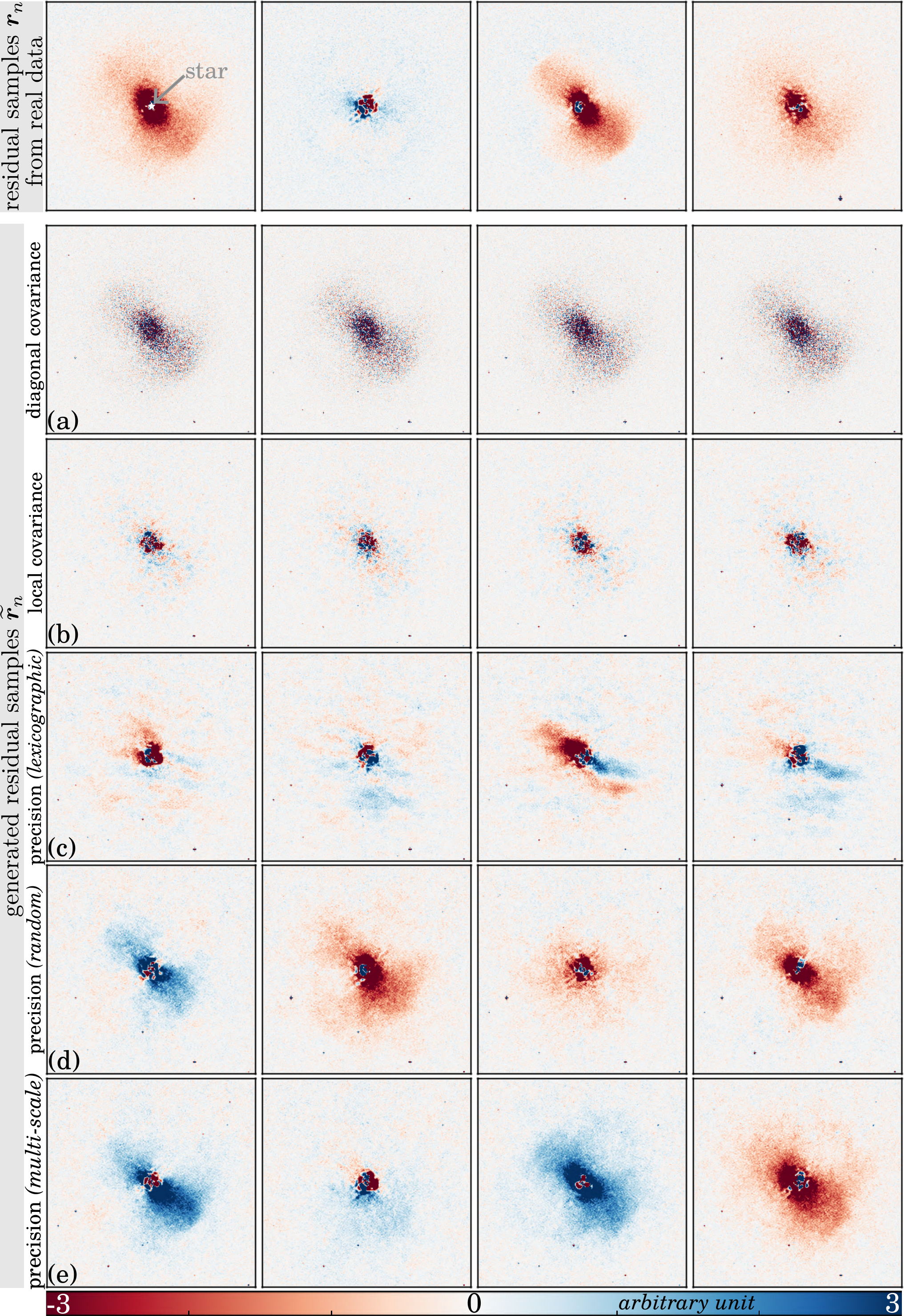}%
	\caption{Top row: examples of real residual samples. Other rows: examples of
          synthetic samples generated from the learned nuisance model. Optimal model
          complexity $M$ has been used for rows (c--e).}
	\label{fig:generated_samples}
\end{figure}

State-of-the-art detection methods exploit this signal diversity
by learning a nuisance model directly from the
data cube $\lbrace \V y_n \rbrace_{n\in\Range{1}{N}}$.
Baseline approaches \cite{VIP} rely on empirical modeling, such as
image combination (\texttt{cADI}) or low-rank approximations via truncated
PCA (\texttt{KLIP}), assuming a
quasi-static nuisance across exposures.
However, as shown in Fig.~\ref{fig:generated_samples}(top), the nuisance is not
strictly stationary across exposures: subtracting the empirical mean
$\estim{\V \mu}$ leaves strong correlated residuals.
More advanced methods, such as \texttt{PACO} \cite{PACO}, use matched filtering and
explicitly model these correlations at a local (patch) scale under a Gaussian
assumption, approximating the nuisance covariance by block-diagonal structures
regularized by shrinkage.
The hybrid \texttt{deep PACO} \cite{dPACO} approach further combines this statistical model with a
deep-learning detector trained in a supervised way.

\subsection{Unsupervised setting and reliability of the nuisance model}
\label{subsec:reliability_nuisance_model}

\begin{figure}[t!]
	\centering
	\includegraphics[width=0.5\textwidth]{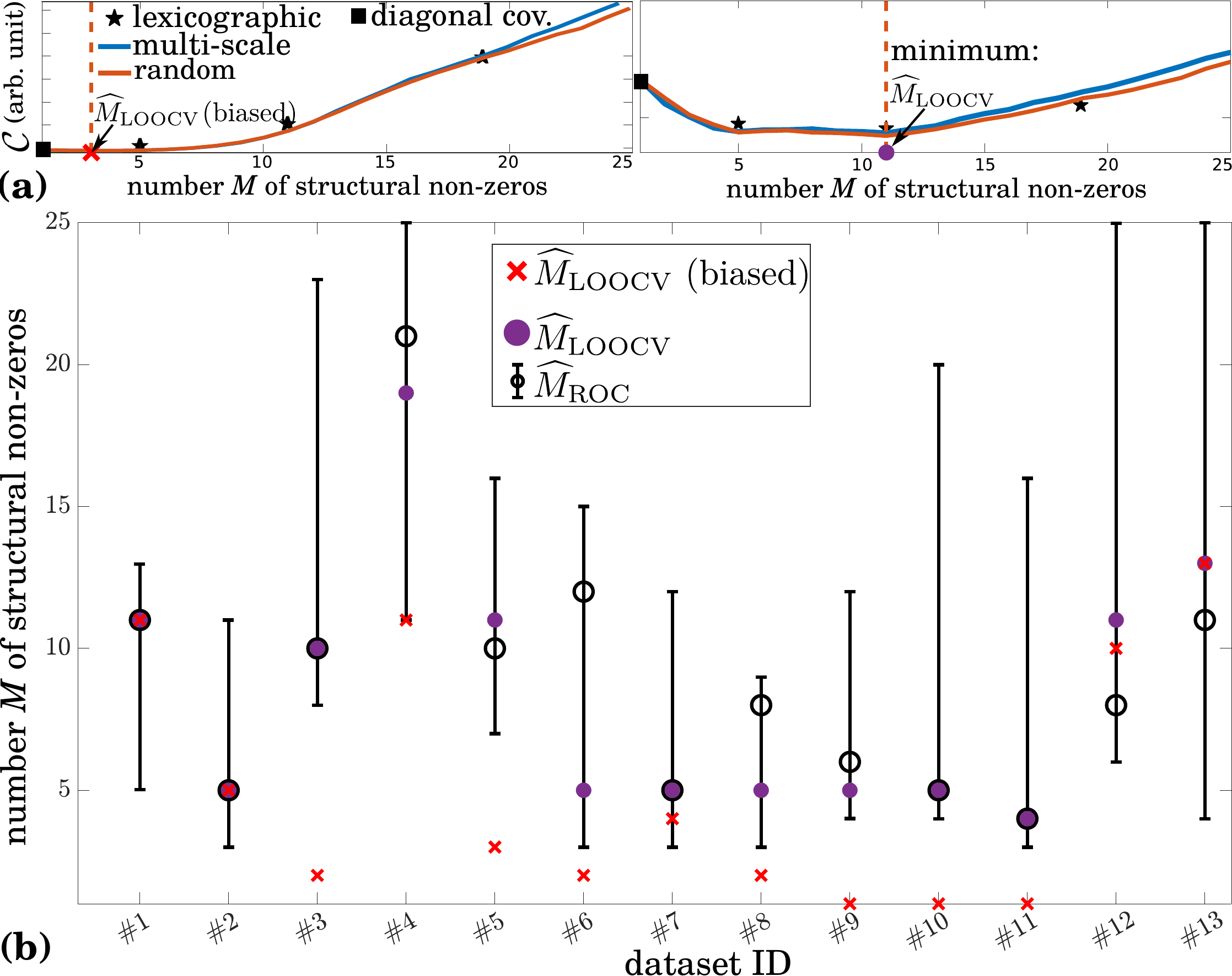}%
	\caption{Unsupervised model complexity selection using LOOCV. (a) LOOCV criterion
          $\mathcal{C}(M)$ without (left) and with (right) unbiasing, for different
          structures of the sparse precision matrix factor $\M R$. (b) Comparison between
          $\estim{M}_{\text{LOOCV}}$ and $\estim{M}_{\text{ROC}}$; error bars indicate model
          complexities yielding detection performance (AUC) within 10\% of the optimum.}
	\label{fig:loocv_summary_combined}
\end{figure}

\noindent We assess the relevance of the proposed unsupervised and unbiased LOOCV criterion
for selecting the nuisance model complexity $M$ by comparing the resulting estimate
$\estim{M}_{\text{LOOCV}}$ with a supervised reference $\estim{M}_{\text{ROC}}$. The latter
is obtained by maximizing the area under the ROC curve (AUC), which summarizes the
precision–recall trade-off via the true positive rate (TPR) and false discovery rate (FDR).
The AUC is computed from extensive simulations involving the injection of synthetic sources
into real data using model~\eqref{eq:direct_model}. These simulations are realistic, as the
nuisance component comes from real observations and the PSF model is accurately calibrated
from instrumental data. Figure~\ref{fig:loocv_summary_combined} shows on 13 datasets that
$\estim{M}_{\text{LOOCV}}$ agrees well with $\estim{M}_{\text{ROC}}$ when LOOCV is
implemented in an unbiased manner (see Sect.~\ref{sec:loocv}). The LOOCV criterion also
enables comparison of sparse structures, with the multi-scale structure generally performing
best. In the following, both the sparse structure and the model complexity are selected
fully unsupervised using LOOCV.


Beyond model selection, we qualitatively assess the expressivity of the proposed Gaussian
nuisance model. Using nuisance statistics $\estim{\M \Omega}$ learned from real data,
Fig.~\ref{fig:generated_samples} compares synthetic residual samples
$\widetilde{\V r}_n = \estim{\M S}\T \V \epsilon_n$ generated by correlating i.i.d. Gaussian
vectors $\V \epsilon_n \in \mathbb{R}^P$ with the learned sparse precision matrix factor
$\M S\T$. Ignoring correlations, or modeling them only locally (local covariance or with the
\textit{lexicographic} structure of the sparse precision matrix factor), fails to reproduce
the large-scale structures observed in real data. In contrast, the \textit{random} and
\textit{multi-scale} structures of the sparse precision matrix factor generate very
realistic samples, demonstrating the ability of the proposed model to capture both local and
long-range correlations.

\subsection{Performance assessment}
\label{subsec:detection_performance}

\begin{figure}[t!]
	\centering
	\includegraphics[width=0.5\textwidth]{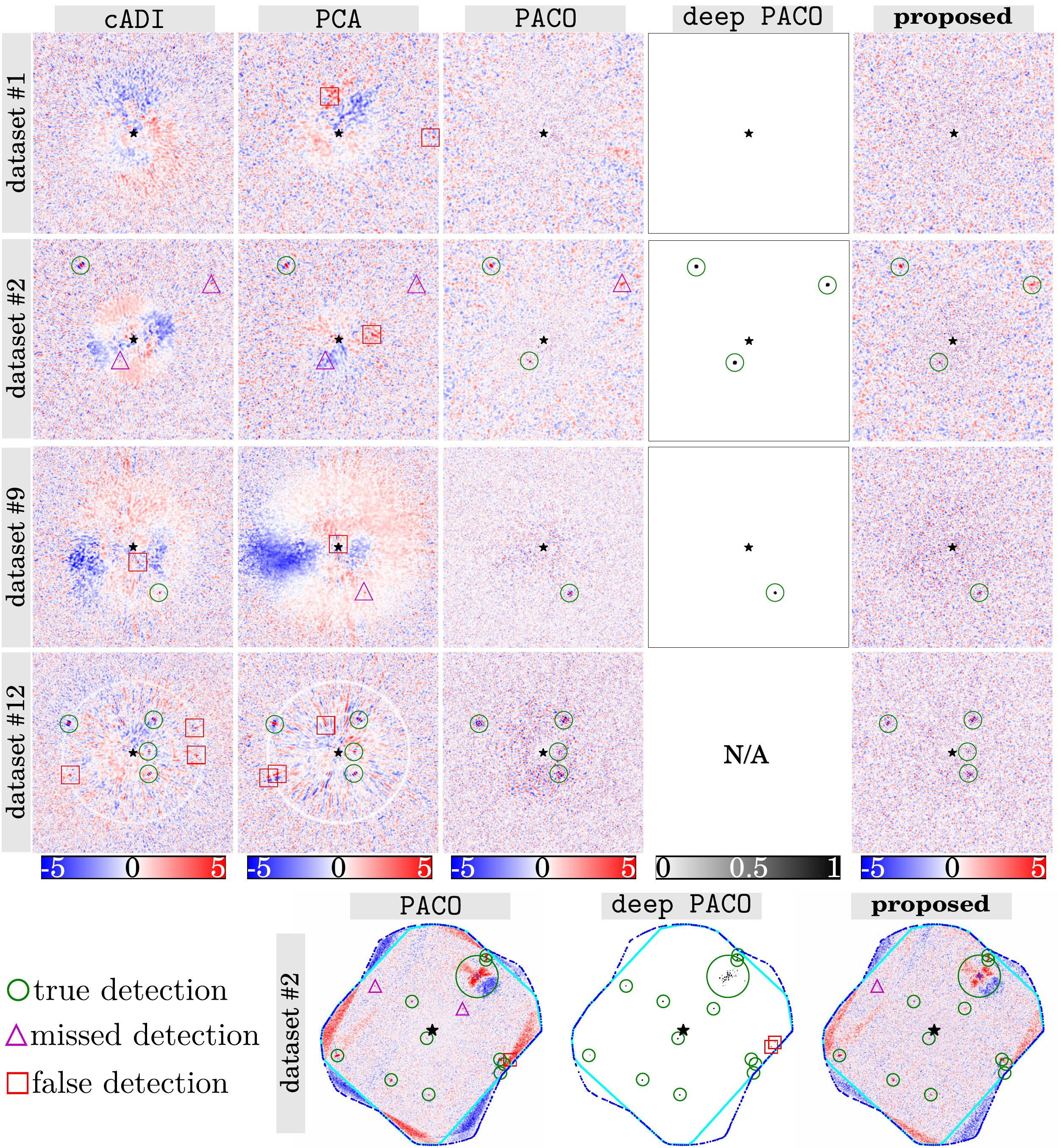}%
	\caption{Detection maps $\Gamma(\V\theta)$ on real data. Top: $200 \times 200$ pixel
          region centered on the star. Bottom: full field of view for dataset \#2, shown for
          the three best-performing algorithms. The detection threshold is set to $\tau=5$
          for \texttt{cADI}, \texttt{PCA}, \texttt{PACO}, and the proposed method (SNR-based
          maps), and to $\tau=0.5$ for \texttt{deep PACO} (pseudo-probability map),
          corresponding to comparable targeted FDRs. \texttt{deep PACO} could not be applied
          to dataset \#12 acquired in a spectral mode inducing highly structured nuisance
          patterns making the training procedure unstable.}
	\label{fig:detection_maps_roi_fullfig}
\end{figure}

\noindent We assess the detection performance of the proposed approach by comparing it with
the state-of-the-art methods described in Sect.~\ref{subsec:hci_principle} on 13 datasets
acquired under various instrumental configurations and observing conditions.
Figure~\ref{fig:detection_maps_roi_fullfig} shows representative detection maps, while
Table~\ref{tab:auc_fdr} reports quantitative results in terms of average AUC and TPR at a
low FDR (0.005), a regime of particular interest in practice. Metrics are computed on
datasets processed by all algorithms for both real sources (72 in total) and synthetic
sources generated as described in Sect.~\ref{subsec:reliability_nuisance_model}, allowing
evaluation over a wide range of fluxes and positions. The proposed method is highly
competitive with respect to state-of-the-art approaches, and often achieves the best
performance in the low-FDR regime which is of particular interest for astronomers.

\begin{table}[t!]
	\caption{Detection metrics on a $200\times200$ pixel ROI centered on the star. Best scores are shown in bold, second best underlined.}
	\begin{center}
		\begin{tabular}{cccccc}
			\toprule
			Metric / Algo. & cADI & PCA & PACO & deep PACO & \textit{proposed}\\
			\midrule
			& \multicolumn{5}{c}{\textit{Real sources}} \\
			AUC & 0.64 & 0.60 & 0.81 & \textbf{0.86} & \underline{0.83}\\
			TPR (FDR=0.005) & 0.53 & 0.50 & \underline{0.75} & \underline{0.75} & \textbf{0.78}\\
			\midrule
			& \multicolumn{5}{c}{\textit{Simulated synthetic sources in real nuisance data}} \\
			AUC	& 0.59 & 0.55 & 0.74 & \textbf{0.90} & \underline{0.89}\\
			TPR (FDR=0.005) & 0.52 & 0.49 & 0.72 & \underline{0.77} & \textbf{0.86}\\
			\bottomrule
		\end{tabular}
		\label{tab:auc_fdr}
	\end{center}
\end{table}
\begin{figure}[t!]
	\centering
	\includegraphics[width=0.5\textwidth]{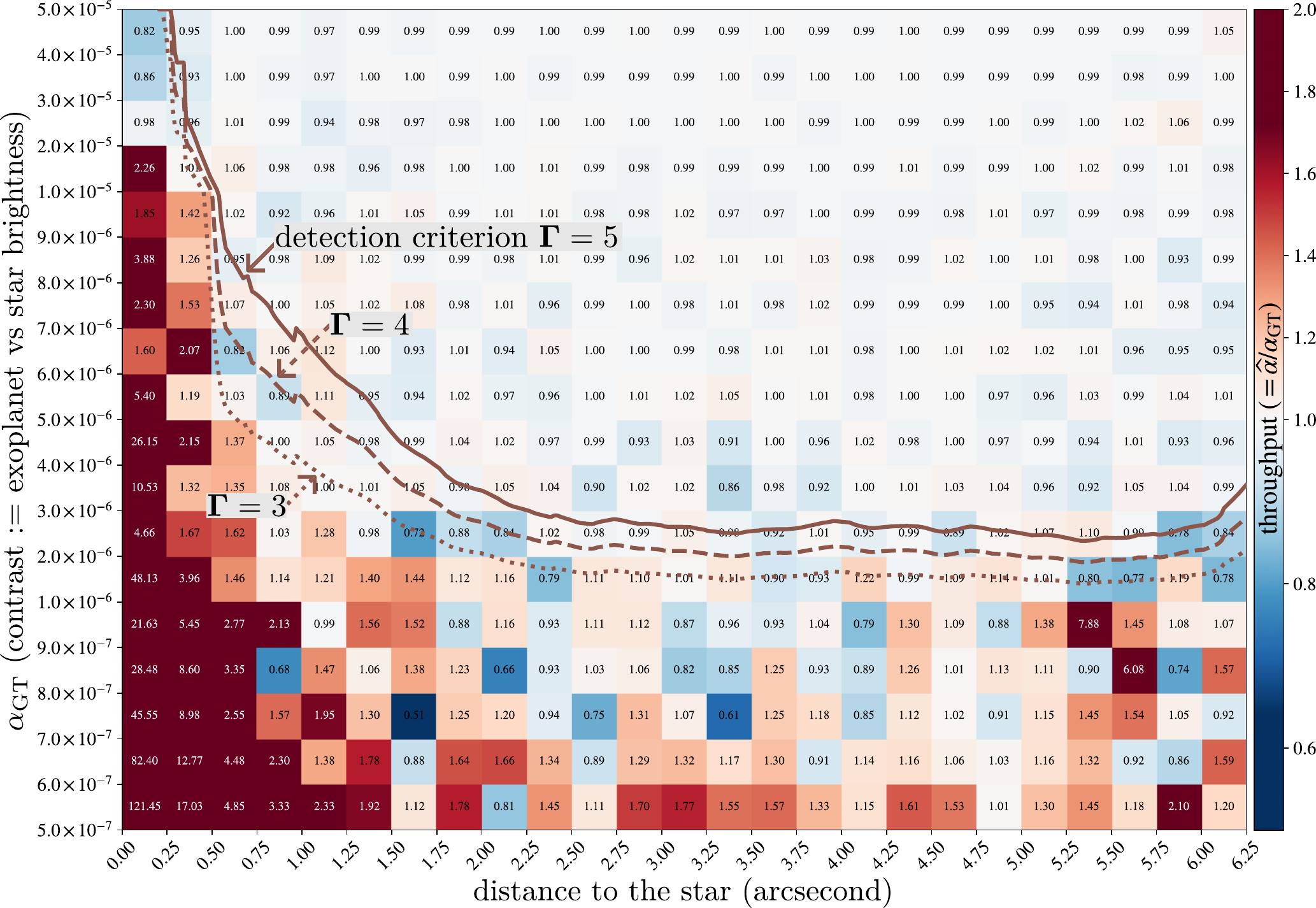}%
	\caption{Estimation metric: color-coded ratio $\widehat{\alpha}/\alpha_{\text{GT}}$ as a function of the
groundtruth source flux $\alpha_{\text{GT}}$ and of the distance to the star. Overlaid curves indicate sensitivity limits at $\Gamma \ge 5$, $4$, and
$3$ (solid, dashed, and dotted). Sources above these contours are detected with
equal or higher confidence.}
	\label{fig:throughput_grid}
\end{figure}
Using the same set of simulated sources, Fig.~\ref{fig:throughput_grid} evaluates flux
estimation accuracy by comparing $\estim{\alpha}$ to the groundtruth
$\alpha_{\text{GT}}$ as a function of separation. The estimates are unbiased and
in excellent agreement for sources detected at SNR $\Gamma \ge 3$. At lower
detection confidence, the estimation degrades, particularly close to the star,
where stronger stellar residuals and reduced
signal diversity induced by the limited apparent rotation of the sources make the unmixing
task more difficult.

\section{Conclusions and Perspectives}
\label{sec:conclusion}

\noindent We proposed a data-driven approach to model correlated multivariate data in the
low-sample regime. A low-complexity model is learned through a sparse
decomposition of the precision matrix, with the model complexity selected
via an unsupervised procedure. Integrating this statistical
model into a detection framework yields a CFAR detector achieving state-of-the-art performance,
particularly at low FDR.
Future work will extend this approach to other imaging modalities and tasks, such
as multi-spectral high-contrast imaging.

\vspace*{2ex}\noindent\begin{minipage}{\columnwidth}
	\footnotesize
\bibliographystyle{IEEEtran}
\bibliography{asap_detection_eusipco_2026}
\end{minipage}


\end{document}